\documentclass[11pt,a4paper]{article}

\usepackage[margin=1in]{geometry}
\usepackage{graphicx}
\usepackage{float}

\usepackage{setspace}
\title{A Security Analysis of Long-Horizon Agentic AI Systems: Threats, Evaluation, and Framework Development}

\author{
Ahmed Mohammed Almalki$^{1}$ \hspace{2cm} Mehedi Masud$^{2}$\\[0.8em]
{\small $^{1,2}$ Department of Computer Science}\\
{\small College of Computers and Information Technology}\\
{\small Taif University, KSA}
}

\date{Summer 2026}

\begin{document}

\maketitle

\begin{abstract}
This paper presents a structured analysis of security challenges in long-horizon agentic AI systems. The study reviews existing threats, evaluation approaches, attack propagation mechanisms, and security frameworks. A taxonomy of security threats and a framework for analyzing attack propagation are proposed to support future research in agentic AI security.
\end{abstract}


\section{Introduction}

\subsection{Background and Motivation}

Agentic Artificial Intelligence (AI) systems represent an evolution beyond traditional Large Language Models (LLMs). While conventional LLMs primarily respond to user prompts, agentic AI systems can autonomously plan, reason, interact with external tools, and perform multi-step tasks to achieve complex objectives.

Recent advances in agentic AI have enabled applications in cybersecurity, healthcare, business automation, and intelligent personal assistants. These systems continuously interact with external environments through APIs, databases, web services, and memory modules, allowing them to operate in dynamic settings.

Long-horizon agentic AI systems are particularly important because they execute tasks across multiple sequential steps. Decisions made during earlier stages may influence future actions, making the system behavior dependent on memory, planning, and historical context.

For example, a cybersecurity agent may continuously monitor network activity, analyze security logs, generate alerts, recommend mitigation strategies, and update future decisions based on previous observations. Such long-horizon workflows create new security challenges that are not adequately addressed by traditional AI security approaches.

The integration of memory, tools, reasoning modules, and external environments significantly expands the attack surface of agentic AI systems. Consequently, security threats may propagate across multiple components and remain effective over extended periods of execution.

\subsection{Problem Statement}

Despite the growing adoption of agentic AI systems, current security research remains fragmented and lacks a unified understanding of security risks in long-horizon environments.

Most existing studies focus on isolated attack categories such as prompt injection, memory poisoning, or tool exploitation. However, these approaches rarely examine how threats propagate across multiple system components, including memory, reasoning modules, planning mechanisms, and external tools.

Furthermore, existing evaluation approaches are often designed for traditional single-step AI systems and fail to capture attack persistence, long-term effects, and multi-step attack propagation.

The absence of a unified taxonomy, structured evaluation methodologies, and comprehensive analytical frameworks creates significant challenges in understanding and mitigating security threats in long-horizon agentic AI systems.

\subsection{Research Objectives}

This study aims to:

\begin{itemize}
\item Review and synthesize existing literature on security in long-horizon agentic AI systems.
\item Identify and categorize major security threats and attack surfaces.
\item Analyze how attacks propagate across memory, tools, planning modules, and multi-agent environments.
\item Develop a taxonomy for organizing security threats.
\item Examine existing evaluation methods and benchmarking approaches.
\item Propose a structured framework for analyzing security risks in agentic AI systems.
\end{itemize}

\subsection{Paper Organization}

The remainder of this paper is organized as follows.

Section 2 introduces the theoretical background of agentic AI systems and their security challenges.

Section 3 describes the research methodology.

Section 4 reviews the existing literature on agentic AI security threats.

Section 5 presents the proposed taxonomy, framework, and structured evaluation approach.

Section 6 provides a comparison of related works and identifies current research gaps.

Section 7 discusses the implications of the findings.

Finally, Section 8 concludes the paper and highlights future research directions.

\section{Theoretical Background}

\subsection{Agentic AI Systems}

Agentic AI systems extend the capabilities of traditional Large Language Models by incorporating autonomous planning, reasoning, memory management, and interaction with external tools, as discussed by Xu [8] and Bandi et al. [9]. Unlike conventional LLMs that primarily generate responses to user prompts, agentic systems actively perform tasks and make decisions to achieve predefined objectives.

These systems can interact with APIs, databases, browsers, software applications, and external environments. As a result, agentic AI systems are increasingly used in domains such as cybersecurity, healthcare, business automation, robotics, and intelligent assistants.

A practical example is an AI travel assistant that autonomously searches for flights, compares hotel options, checks weather forecasts, updates calendars, and completes bookings through external services. This demonstrates how agentic systems perform coordinated multi-step actions rather than isolated responses.

\subsection{Long-Horizon Agentic AI Systems}

Long-horizon agentic AI systems execute tasks across multiple sequential steps where each action depends on previous outcomes \cite{ref1,ref2}. These systems maintain context through memory and continuously update their behavior as new information becomes available.

The long-horizon nature of these systems introduces dependencies between earlier and later decisions. Consequently, errors, malicious instructions, or manipulated information introduced at one stage may influence future actions and overall system behavior.

A representative example is a cybersecurity monitoring agent that continuously collects logs, analyzes suspicious activities, generates alerts, recommends mitigation actions, and updates future decisions based on previously observed events. Such workflows require continuous reasoning, planning, and memory utilization across extended periods.

Long-horizon agentic AI systems are increasingly deployed in cybersecurity monitoring, healthcare decision support, autonomous software engineering, business process automation, and intelligent personal assistants.

\subsection{Security Risks and Attack Surfaces}

The integration of memory systems, reasoning modules, external tools, and dynamic environments significantly increases the attack surface of agentic AI systems.

Security threats may enter through various attack surfaces including:

\begin{itemize}
\item Input Channels
\item Reasoning Modules
\item Memory Systems
\item Tool Interfaces
\item External Environments
\end{itemize}

Input-channel attacks often involve malicious instructions embedded in user prompts, retrieved documents, or web content. Reasoning-module attacks attempt to manipulate intermediate decision-making processes.

Memory-based attacks target stored information and may introduce persistent security risks that affect future decisions. Tool-related attacks exploit APIs, databases, and external services used by the agent.

External-environment attacks manipulate information sources that agents rely upon during task execution.

These attack surfaces create opportunities for attack propagation, persistence, unauthorized actions, data leakage, and system manipulation in long-horizon environments.
\begin{table}[H]
\centering
\small
\begin{tabular}{|p{2.8cm}|p{5cm}|p{4.5cm}|}
\hline
\textbf{Attack Surface} & \textbf{Example Scenario} & \textbf{Potential Impact} \\
\hline

Input Channels &
A malicious webpage contains hidden prompt injection instructions that override the agent's original task. &
Unauthorized actions, information leakage, and task manipulation. \\
\hline

Reasoning Modules &
An attacker manipulates intermediate reasoning steps, causing the agent to draw incorrect conclusions. &
Faulty decision-making and deviation from intended objectives. \\
\hline

Memory Systems &
False information is injected into long-term memory and later reused as trusted knowledge. &
Persistent errors, misinformation propagation, and incorrect future decisions. \\
\hline

Tool Interfaces &
A compromised API returns manipulated outputs that are accepted by the agent as legitimate results. &
Execution of unsafe actions, incorrect recommendations, and system misuse. \\
\hline

External Environments &
The agent retrieves poisoned data from external sources and incorporates it into its workflow. &
Incorrect reasoning, unreliable outputs, and attack propagation across multiple tasks. \\
\hline

\end{tabular}
\caption{Example Attack Scenarios and Their Potential Impact Across Different Attack Surfaces}
\label{tab:attack_scenarios}
\end{table}
Table~\ref{tab:attack_scenarios} presents representative attack scenarios for major attack surfaces in long-horizon agentic AI systems and highlights their potential security impact.

\section{Methodology}

\subsection{Review Design}

This study adopts a survey-based research methodology to analyze security challenges in long-horizon agentic AI systems. The research focuses on reviewing, synthesizing, and comparing existing studies related to security threats, attack propagation, defense mechanisms, and evaluation approaches.

The study follows a structured analytical process that includes threat identification, taxonomy development, framework construction, and comparative analysis of existing literature.

\subsection{Data Sources and Paper Selection}

The study is based on a collection of recent research papers related to agentic AI security, large language model security, prompt injection attacks, memory poisoning, tool exploitation, multi-agent security, and evaluation methodologies.

A total of twenty recent academic papers were selected based on their relevance to the research topic, contribution to agentic AI security, and coverage of long-horizon attack scenarios.

The selected studies include survey papers, security assessments, attack analyses, benchmarking studies, and framework-oriented research.

\subsection{Analysis Method}

The analysis consists of four main stages.

\begin{enumerate}

\item Literature Review and Synthesis

Relevant studies are reviewed to identify major security threats, attack surfaces, defense mechanisms, and evaluation approaches.

\item Taxonomy Construction

Security threats are categorized according to attack entry points, propagation mechanisms, and affected system components.

\item Framework Development

A conceptual framework is developed to illustrate how attacks propagate through memory modules, planning mechanisms, external tools, and output layers.

\item Comparative Analysis

Existing studies are compared based on attack coverage, long-horizon analysis, consideration of memory and tools, and evaluation methodologies.

\end{enumerate}

This methodology provides a systematic approach for understanding emerging security challenges in long-horizon agentic AI systems.

\section{Literature Review}

\subsection{Prompt Injection and Input-Based Attacks}

Prompt injection is one of the most significant security threats in agentic AI systems, as highlighted by Wang et al. [4] and Liu et al. [13]. In this attack, malicious instructions are embedded within external inputs such as webpages, retrieved documents, or tool outputs. The agent may interpret these instructions as legitimate commands, resulting in unintended actions.

Research has shown that prompt injection attacks become particularly dangerous in long-horizon environments because malicious instructions may persist across multiple interaction steps and influence future decisions. Existing defenses such as input filtering and prompt hardening provide partial protection but often fail to address delayed attack activation and attack propagation.

A typical attack scenario involves a malicious webpage containing hidden instructions that cause the agent to ignore its original objectives and execute attacker-controlled actions.

\subsection{Memory-Based Attacks}

Memory systems enable agentic AI systems to maintain context and continuity across multiple interactions. However, these capabilities introduce new attack opportunities.

Memory poisoning attacks occur when adversaries inject false or manipulated information into the system's memory, as discussed by Fendley et al. [10] and Souly et al. [15]. Since future decisions may rely on stored information, poisoned memory can influence behavior over extended periods.

Research indicates that memory-based attacks represent one of the most challenging threats in long-horizon environments because their effects may remain hidden until future stages of execution.

\subsection{Tool Exploitation and Tool-Chain Attacks}

Agentic AI systems frequently interact with external tools such as APIs, databases, search engines, and software applications \cite{ref12,ref18}.

Attackers may exploit vulnerabilities within these tools or manipulate their outputs to influence agent behavior. Tool-chain attacks can trigger cascading effects where compromised outputs lead to additional unsafe actions.

The increasing reliance on external services expands the attack surface and introduces risks that cannot be mitigated solely through input-level defenses.

\subsection{Multi-Agent and Trust-Based Attacks}

Many modern agentic AI environments involve multiple collaborating agents \cite{ref17}.

Although cooperation improves efficiency and scalability, it also introduces trust-related vulnerabilities. Malicious agents may distribute manipulated information, influence coordination processes, or exploit trust assumptions among collaborating agents.

Research highlights the need for trust verification mechanisms and secure communication protocols to reduce risks in multi-agent environments.

\subsection{Planning and Goal-Manipulation Attacks}

Planning attacks target the reasoning and decision-making capabilities of agentic AI systems \cite{ref14,ref20}.

Adversaries may manipulate goals, sub-goals, or intermediate reasoning steps to redirect system behavior toward unintended objectives. Such attacks are especially dangerous in long-horizon environments because compromised planning decisions may affect multiple future actions.

Goal hijacking represents one of the most severe outcomes of planning attacks, potentially causing the system to deviate significantly from its intended objectives.

\subsection{Evaluation and Benchmarking}

Evaluating security in agentic AI systems remains a significant challenge \cite{ref1,ref19}.

Traditional evaluation approaches focus on single-step interactions and therefore fail to capture attack persistence, attack propagation, and long-term behavioral changes.

Recent studies emphasize the importance of prompt-level evaluation, task-level assessment, long-horizon simulations, and structured benchmarking methodologies. However, standardized evaluation frameworks for agentic AI security remain limited.

This gap motivates the development of more comprehensive evaluation approaches capable of measuring security performance across extended execution horizons.

\section{Proposed Taxonomy and Framework}

\subsection{Proposed Taxonomy}

To provide a structured understanding of security threats in long-horizon agentic AI systems, this study proposes a taxonomy based on attack entry points, propagation mechanisms, and affected system components.

The taxonomy categorizes threats into five major groups:

\begin{itemize}
\item Input-Based Attacks
\item Memory Attacks
\item Tool-Related Attacks
\item Planning Attacks
\item Multi-Agent Attacks
\end{itemize}

Input-based attacks enter the system through prompts, retrieved documents, webpages, or external content. These attacks attempt to manipulate agent behavior at the point of entry.

Memory attacks target stored information and seek to introduce malicious or misleading content that may influence future decisions.

Tool-related attacks exploit APIs, databases, and external services used by the agent. Manipulated tool outputs may affect decision-making and system actions.

Planning attacks target reasoning processes, goals, and intermediate decision-making steps.

Multi-agent attacks exploit communication channels and trust relationships between collaborating agents.

Together, these categories provide a structured view of security risks in long-horizon agentic AI systems and support systematic threat analysis.
\begin{figure}[H]
\centering
\includegraphics[width=\textwidth]{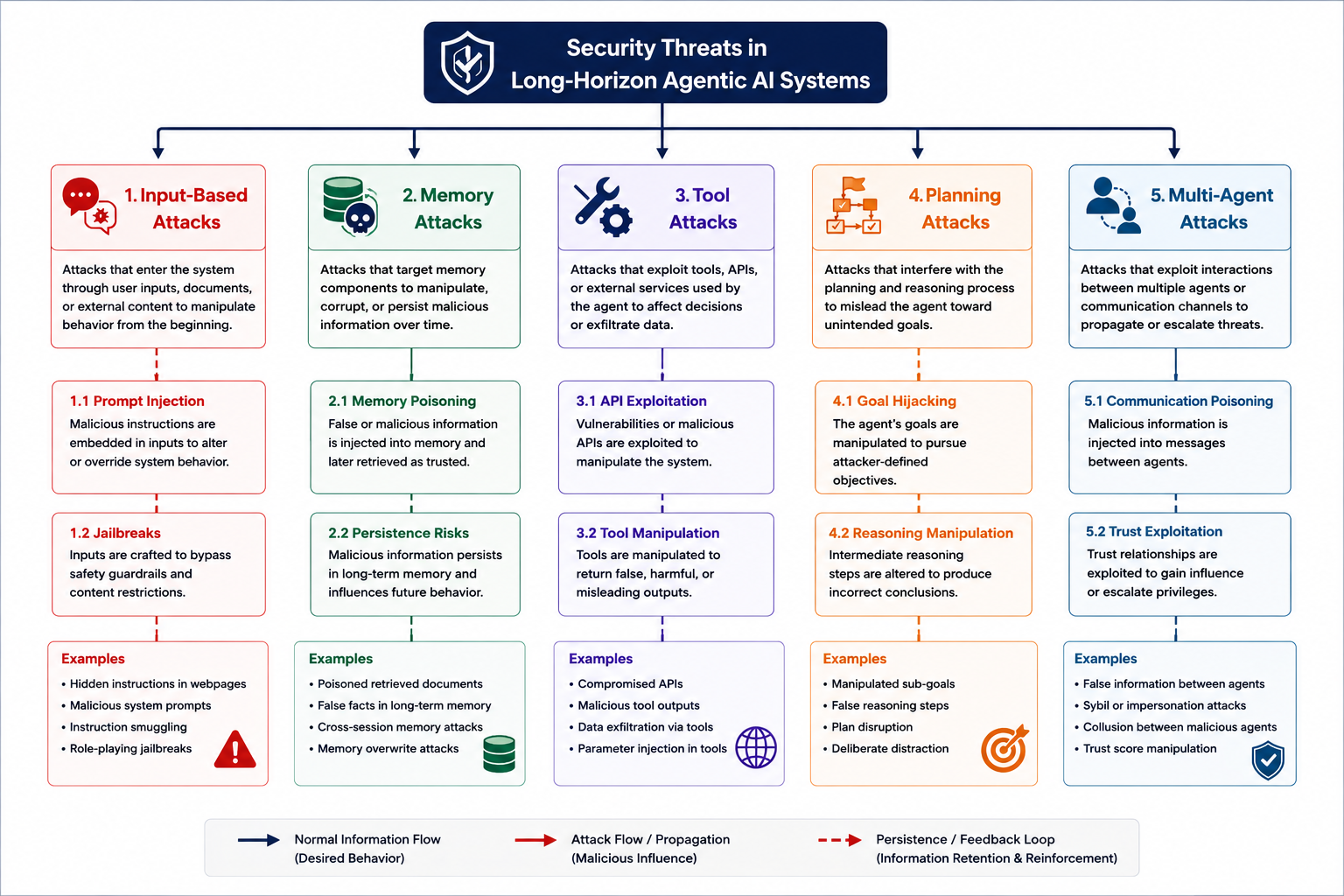}
\caption{Proposed Taxonomy of Security Threats in Long-Horizon Agentic AI Systems}
\label{fig:taxonomy}
\end{figure}
Figure~\ref{fig:taxonomy} illustrates the proposed taxonomy of security threats in long-horizon agentic AI systems.

\subsection{General Framework}

This study proposes a conceptual framework for analyzing how attacks propagate through long-horizon agentic AI systems.

The framework consists of five primary layers:

\begin{enumerate}
\item Input Layer
\item Agent Core
\item Memory Module
\item Tool Layer
\item Output Layer
\end{enumerate}

Security threats may enter through external inputs and subsequently propagate through memory systems, planning mechanisms, and tool interactions before influencing final outputs.

The framework highlights how attack propagation, persistence, and component interaction contribute to security risks in long-horizon environments.

\begin{figure}[H]
\centering
\includegraphics[width=\textwidth]{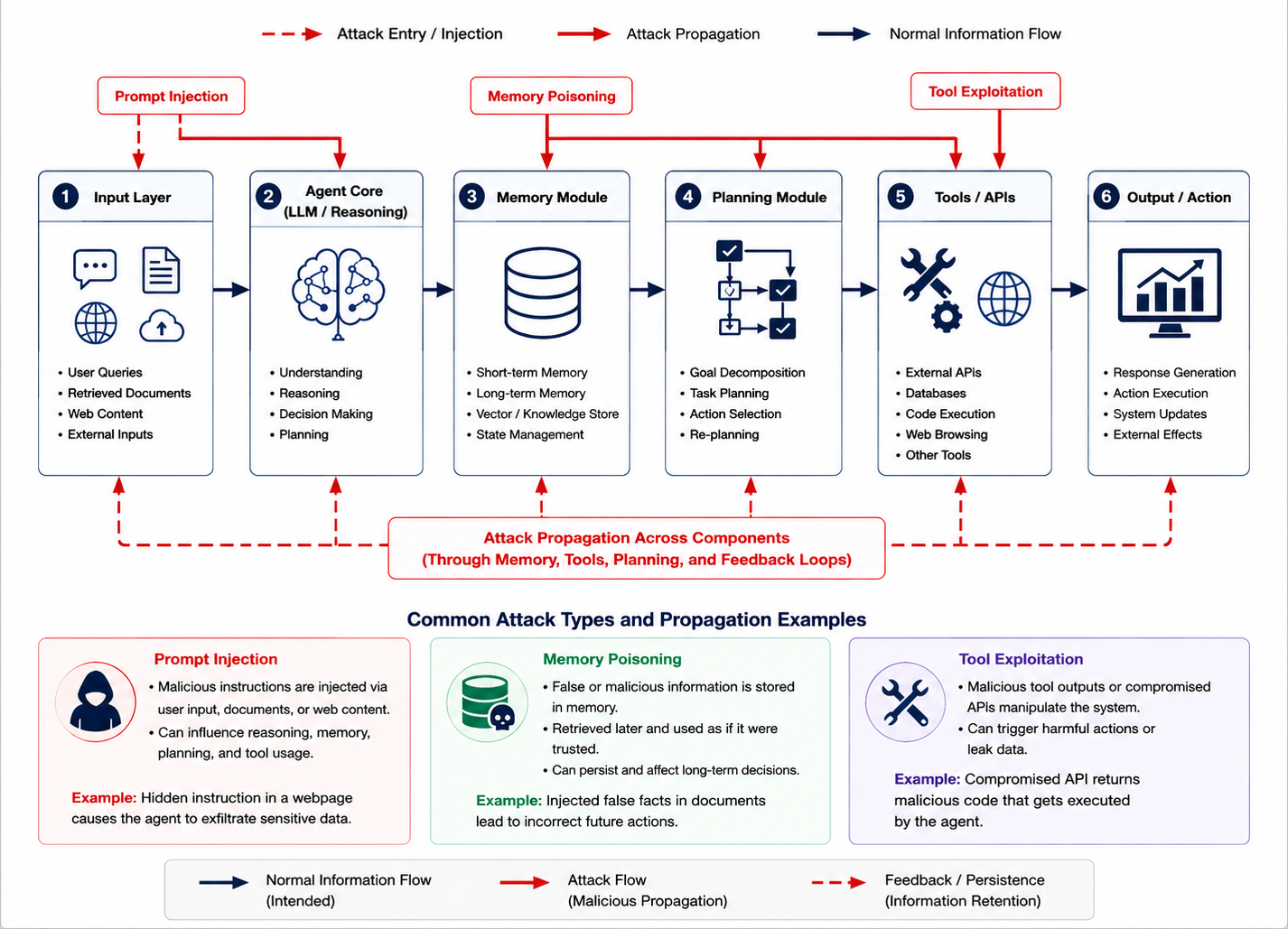}
\caption{Proposed Security Analysis Framework for Long-Horizon Agentic AI Systems}
\label{fig:framework}
\end{figure}

Figure~\ref{fig:framework} presents the proposed framework for analyzing attack propagation and security risks.

\subsection{Structured Evaluation Approach}

To support systematic security assessment, this study proposes a structured evaluation approach based on four dimensions:

\begin{enumerate}

\item Multi-Step Execution Analysis

Evaluates how attacks influence system behavior across multiple interaction steps.

\item Attack Propagation Assessment

Measures how malicious information spreads through memory, reasoning modules, tools, and outputs.

\item Persistence Measurement

Assesses whether attacks remain effective over time and continue influencing future decisions.

\item System Response Evaluation

Evaluates the system's ability to detect, mitigate, and recover from security threats.

\end{enumerate}

These dimensions provide a consistent framework for evaluating security risks in long-horizon agentic AI systems.

\section{Comparison and Research Gaps}

\subsection{Comparison of Related Works}

\begin{table}[H]
\centering
\small
\begin{tabular}{|p{3cm}|p{2cm}|p{2cm}|p{2cm}|}
\hline
Study & Long-Horizon Analysis & Attack Coverage & Evaluation Method \\
\hline
Prompt Injection Studies & Limited & Input Attacks & Basic Testing \\
\hline
Memory Security Studies & Partial & Memory Attacks & Experimental \\
\hline
Tool Exploitation Studies & Partial & Tool Attacks & Case-Based \\
\hline
Multi-Agent Security Studies & Partial & Trust Attacks & Simulation \\
\hline
Benchmarking Studies & Limited & Multiple Threats & Benchmark Evaluation \\
\hline
Proposed Work & Comprehensive & Multiple Attack Surfaces & Structured Evaluation Framework \\
\hline
\end{tabular}
\caption{Comparison Between Existing Studies and the Proposed Work}
\label{tab:comparison}
\end{table}

\subsection{Research Gaps}

The literature reveals several important research gaps.

\begin{itemize}

\item Lack of unified security taxonomies for long-horizon agentic AI systems.

\item Limited evaluation methodologies capable of measuring attack persistence and propagation.

\item Fragmented analysis of memory, tools, planning modules, and multi-agent environments.

\item Insufficient framework-oriented approaches for analyzing complex attack interactions.

\item Lack of standardized benchmarking procedures for agentic AI security.

\end{itemize}

Addressing these gaps is essential for improving the security, reliability, and trustworthiness of future agentic AI systems.

\section{Discussion}

The rapid evolution of agentic AI systems introduces security challenges that differ significantly from those observed in traditional AI environments.

The combination of memory, planning, tool usage, and multi-step execution creates opportunities for attack propagation and persistence. Existing security approaches remain largely focused on isolated threats and therefore may not adequately address long-horizon risks.

The proposed taxonomy, framework, and evaluation approach provide a structured foundation for future research in agentic AI security.

\section{Conclusion}

This paper presented a structured analysis of security challenges in long-horizon agentic AI systems.

The study reviewed major threat categories, examined attack propagation mechanisms, analyzed current evaluation approaches, and identified key research gaps.

In addition, a security taxonomy, conceptual framework, and structured evaluation methodology were proposed to support future research and security assessment efforts.

As agentic AI systems continue to evolve, developing comprehensive security frameworks will become increasingly important for ensuring reliability, safety, and trustworthiness.


\begin{thebibliography}{99}

\bibitem{ref1}
Chhabra, A., Datta, S., Nahin, S. K., \& Mohapatra, P. (2026).
Agentic AI Security: Threats, Defenses, Evaluation, and Open Challenges.

\bibitem{ref2}
Lazer, S. J., Aryal, K., Gupta, M., \& Bertino, E. (2026).
A Survey of Agentic AI and Cybersecurity: Challenges, Opportunities and Use-case Prototypes.

\bibitem{ref3}
Grimes, K., Lawler, J., Garrett, R. C., et al. (2025).
SOK: Bridging Research and Practice in LLM Agent Security.

\bibitem{ref4}
Wang, P., Li, X., Xiang, C., et al. (2025).
The Landscape of Prompt Injection Threats in LLM Agents: From Taxonomy to Analysis.

\bibitem{ref5}
Leo, A., et al. (2026).
From Threat to Trust: Assessing Security Risks of Agentic AI Systems.

\bibitem{ref6}
Huang, X., Karthick, V. B., Chen, T., et al. (2025).
Trust in LLM-controlled Robotics: A Survey of Security Threats, Defenses and Challenges.

\bibitem{ref7}
Schwarz, D. (2025).
Countermind: A Multi-Layered Security Architecture for Large Language Models.

\bibitem{ref8}
Xu, B. (2026).
AI Agent Systems: Architectures, Applications, and Evaluation.

\bibitem{ref9}
Bandi, A., Kongari, B., Naguru, R., et al. (2025).
The Rise of Agentic AI: A Review of Definitions, Frameworks, Architectures, Applications, Evaluation Metrics, and Challenges.

\bibitem{ref10}
Fendley, N., Staley, E. W., Carney, J., et al. (2025).
A Systematic Review of Poisoning Attacks Against Large Language Models.

\bibitem{ref11}
Song, S. (2024).
A Survey of Textual Adversarial Attacks and Defenses on Large Language Models.

\bibitem{ref12}
Xie, Y., Luo, M., Liu, Z., et al. (2025).
Red-Teaming Coding Agents from a Tool-Invocation Perspective: An Empirical Security Assessment.

\bibitem{ref13}
Liu, Y., Jia, Y., Geng, R., et al. (2024).
Formalizing and Benchmarking Prompt Injection Attacks and Defenses.

\bibitem{ref14}
Lupinacci, M., Pironti, F. A., Blefari, F., et al. (2025).
The Dark Side of LLMs: Agent-based Attacks for Complete Computer Takeover.

\bibitem{ref15}
Souly, A., Rando, J., Chapman, E., et al. (2025).
Poisoning Attacks on LLMs Require a Near-Constant Number of Poison Samples.

\bibitem{ref16}
Gao, Y., Shumailov, I., \& Fawaz, K. (2025).
Supply-Chain Attacks in Machine Learning Frameworks.

\bibitem{ref17}
Standen, M., Kim, J., \& Szabo, C. (2025).
Adversarial Machine Learning Attacks and Defences in Multi-Agent Reinforcement Learning.

\bibitem{ref18}
Sneh, J., Yan, R., Yu, J., et al. (2025).
ToolTweak: An Attack on Tool Selection in LLM-Based Agents.

\bibitem{ref19}
Chen, S., Li, X., Zhang, M., et al. (2025).
CARES: Comprehensive Evaluation of Safety and Adversarial Robustness in Medical LLMs.

\bibitem{ref20}
Wang, X., Zhang, Y., Gong, Z., et al. (2026).
From Helpfulness to Toxic Proactivity: Diagnosing Behavioral Misalignment in LLM Agents.

\end{thebibliography}
\end{document}